\documentclass[11pt,english]{article}
\usepackage{amsmath,amssymb,color,slashed,graphicx,babel}

\makeatletter
\@ifundefined{showcaptionsetup}{}{%
\PassOptionsToPackage{caption=false}{subfig}}
\usepackage{subfig}
\makeatother

\usepackage{amsfonts}

\newcommand{\hoch}[1]{$\, ^{#1}$}

\newcommand{\be}{\begin{equation}}
\newcommand{\ee}{\end{equation}}
\newcommand{\bea}{\setlength\arraycolsep{2pt} \begin{eqnarray}}
\newcommand{\eea}{\end{eqnarray}}

\def\ft#1#2{{\textstyle{\frac{\scriptstyle #1}{\scriptstyle #2} } }}
\def\fft#1#2{{\frac{#1}{#2}}}

\def\0{{\sst{(0)}}}
\def\1{{\sst{(1)}}}
\def\2{{\sst{(2)}}}
\def\3{{\sst{(3)}}}
\def\4{{\sst{(4)}}}
\def\5{{\sst{(5)}}}
\def\6{{\sst{(6)}}}
\def\7{{\sst{(7)}}}
\def\8{{\sst{(8)}}}
\def\sst#1{{\scriptscriptstyle #1}}

\begin{document}


\vspace{25pt}
\begin{center}
{\large {\bf Solutions of Free Higher Spins in AdS}}

\vspace{10pt}

H. L\"u\hoch{1,2} and Kai-Nan Shao\hoch{3}

\vspace{10pt}

\hoch{1}{\it China Economics and Management Academy\\
Central University of Finance and Economics, Beijing 100081}

\vspace{10pt}

\hoch{2}{\it Institute for Advanced Study, Shenzhen University\\
Nanhai Ave 3688, Shenzhen 518060}

\vspace{10pt}

\hoch{3}{\it Zheijiang Institute of Modern Physics\\
Department of Physics, Zheijiang University, Hangzhou 310027}

\vspace{40pt}

\underline{ABSTRACT}
\end{center}

We consider free massive and massless higher integer spins in
AdS backgrounds in general $D$ dimensions.  We obtain the solutions
corresponding to the highest-weight state of the spin-$\ell$
representations of the $SO(2,D-1)$ isometry groups.  The solution for the spin-$\ell$ field is expressed recursively in terms of that for the spin-$(\ell-1)$. Thus starting from the explicit spin-0, all the higher-spin solutions can be obtained. These solutions allow us to derive the generalized Breitenlohner-Freedman bound, and analyze the asymptotic falloffs.  In particular, solutions with negative mass square in general have falloffs slower than those of the Schwarzschild AdS black holes in the AdS boundaries.

\vspace{15pt}

\thispagestyle{empty}





\newpage

There has been continuing interest in higher spins since the early
days of quantum field theory.  The non-interacting integer
higher-spin fields can be defined by the Fierz-Pauli conditions
\cite{fp}, namely the Klein-Gordan equation
\begin{equation}
\Big(\Box - (M^{(\ell)})^2\Big)\phi_{\mu_1\cdots\mu_\ell}^{(\ell)}
=0\,,\label{flateom}
\end{equation}
together with the transverse and traceless conditions
\begin{equation}
\nabla^{\mu_1}\phi^{(\ell)}_{\mu_1\cdots\mu_\ell}=0\,,\qquad
g^{\mu_1\mu_2} \phi^{(\ell)}_{\mu_1\cdots\mu_\ell}=0\,,\label{tt}
\end{equation}
where $\phi^{(\ell)}_{\mu_1\cdots \mu_\ell}$ are totally symmetric.
The Lagrangian formulations for the free massive and massless
higher-spin fields were obtained in \cite{sh} and \cite{fron}
respectively.

     It is natural to consider higher-spin gravity theory, which
turns out to be inconsistent in Einstein gravity, which admits the
Minkowski vacuum \cite{ad}. However, when a negative cosmological
constant is introduced with an anti-de Sitter (AdS) vacuum, the
interacting theory of gravity and higher spins can be consistently
defined, at least at the level of equations of motion \cite{vasi1}.
Considerable interest has since been paid to the subject. The
importance of higher-spin gravities is apparent in the context of
string theory in the tensionless limit where the infinite tower of
higher-spin string excitations become massless. From the point of
view of the AdS/CFT correspondence, gravities coupled to higher
spins are expected to be the gravitational dual of the
weakly-coupled conformal field theories. Here we give a few
references \cite{vasi2}-\cite{Bianchi:2004xi} and reviews
\cite{Sorokin:2004ie}-\cite{Bianchi:2005yh}, which are by no means
all inclusive.

When the theory has a cosmological constant, the definition of the
mass in the Klein-Gordan equation is shifted by a constant
$(M^{(\ell)}_{\rm AdS})^2$ that depends on the dimension $D$, spin
$\ell$ and the cosmological constant.  The Klein-Gordan equation
becomes
\begin{equation}
\Big(\Box - (M^{(\ell)}_{\rm AdS})^2 -
(M^{(\ell)})^2\Big)\phi_{\mu_1\cdots\mu_\ell}^{(\ell)}
=0\,,\label{adseom}
\end{equation}
with the same transverse and traceless conditions (\ref{tt}).  In the
massless limit $M^{(\ell)}=0$, there are additional gauge symmetries.  The gauge invariant equation of motion is given by \cite{Bianchi:2005ze}
\begin{eqnarray}
&&\Box \phi^{(\ell)}_{\mu_1\cdots\mu_s} - \ell \nabla_{(\mu_1}
(\nabla\cdot \phi^{(\ell)})_{\mu_2\cdots\mu_\ell)} +
\ft12\ell(\ell-1) \nabla_{(\mu_1}\nabla_{\mu_2}
\phi^{(\ell)}_{\mu_3\cdots \mu_\ell)\lambda}{}^\lambda \cr 
&& - (M^{(\ell)}_{\rm AdS})^2\, \phi^{(\ell)}_{\mu_1\cdots\mu_\ell}
- (\widetilde M^{(\ell)}_{\rm AdS})^2\,
g_{(\mu_1\mu_2}\phi^{(\ell)}_{\mu_3\cdots\mu_\ell\lambda)}
{}^{\lambda}=0\,,\label{adsmassless}
\end{eqnarray}
where
\begin{equation}
(M_{\rm AdS}^{(\ell)})^2 =
\fft{(\ell-2)(D-1)+(\ell-1)(\ell-4)}{L^2}\,,\qquad (\widetilde
M_{\rm AdS}^{(\ell)})^2 = \fft{\ell(\ell-1)}{L^2}\,.
\end{equation}
Here $L$ denotes the the ``radius'' of the AdS space-time.
The gauge transformation rule is given by
\begin{equation}
\delta \phi^{(\ell)}_{\mu_1\cdots\mu_\ell} = \ell\, \nabla_{(\mu_1}
\epsilon_{\mu_2\cdots\mu_\ell)}\,,
\end{equation}
where $\epsilon$ is totally symmetric and traceless. For $\ell=0$, the transverse and traceless conditions do not apply; for $\ell=1$, only the transverse condition is relevant.
Interestingly, as we show in the appendix, for $\ell=2$ and $3$, we
can impose some appropriate gauge fixing such that the transverse and traceless conditions come out naturally. However, for $\ell\ge4$, the conditions (\ref{tt}) arise in a more subtle manner.

    In this paper we are constructing explicit solutions of
(\ref{adseom}) with (\ref{tt}) imposed in arbitrary dimensions for arbitrary
spin $\ell$. The $\ell\ge 1$ solutions form finite representations under the $SO(2,D-1)$ isometry group of AdS$_D$.  To obtain the full spin-$\ell$ representation, we can start with the solution associated with the
highest-weight state, which is defined to be the eigenstate of the
Cartan generators, annihilated by all the positive-root generators.
Once such a solution is constructed, the full
representation can be generated by acting repetitively on it with all possible negative-root generators. This method was
first used to construct explicit graviton modes in three-dimensional
topologically massive gravity \cite{Li:2008dq}. Massless and massive
graviton modes in higher-curvature extended gravities in four
\cite{Bergshoeff:2011ri} and general dimensions \cite{cls} were also
constructed.  The aim of this paper is to employ this technique
to obtain explicit solutions for free higher spins in general dimensions.

    To construct explicit solutions, it is necessary to select
appropriate coordinates.  We shall write the the AdS$_D$ metric in
global coordinates.  The metric for the foliating sphere is written
in such a way that the $U(1)$ coordinates of the corresponding
Cartan generators of the isometry group are manifest.  Thus we
choose the following AdS metric:
\begin{equation}
ds^2= L^2 \Big(-\cosh^2\rho\, d\tau^2 + d\rho^2 + \sinh^2\rho\,
d\Omega_{D-2}^2\Big)\,.
\end{equation}
For odd $D=2n+1$ dimensions, the foliating $S^{D-2}$ is given
by
\begin{eqnarray}
d\Omega_{D-2}^2 &=&d\theta_{1}^{2}
+\cos\theta_{1}d\phi_{n}^{2}+\sin^{2}\theta_{1}
\Bigl(d\theta_{2}^{2}+\cos^{2}\theta_{2}d\phi_{n-1}^{2}\cr &&
+\sin^{2}\theta_{2}\bigl(\cdots(d\theta_{n-1}^{2}+
\cos\theta_{n-1}d\phi_{2}^{2}+\sin\theta_{n-1}
d\phi_{1}^{2})\bigr)\Bigr)\,.
\end{eqnarray}
For even $D=2n$ dimensions, the metric takes the same form but
with $\phi_n=0$. The full set of the $SO(2,D-1)$ Killing vectors for these
AdS metrics were discussed in \cite{cls}.  In particular, the Cartan
generators can be conveniently chosen as
\begin{equation}
H_0={\rm i} \fft{\partial}{\partial\tau}\,,\qquad H_i={\rm i}
\fft{\partial}{\partial \phi_{i}}\,,\qquad i=1,2,\cdots,
[\ft{D-1}{2}]\,.
\end{equation}
Consequently, the positive and negative root generators $(E_{\vec
\alpha_x}, E_{-\vec \alpha_x})$ can all be specified uniquely.  These generators satisfy the $so(2,D-1)$ algebra, namely
\begin{eqnarray}
[H_i,H_j]&=&0\,,\qquad [H_i, E_{\vec a_x}] = \vec \alpha_x^i E_{\vec \alpha_x}\,,\cr
[E_{-\vec\alpha_x}, E_{\vec \alpha_x}] &=& \fft{2}{|\vec \alpha_x|^2} \vec \alpha^i_x H_i\,,\qquad i=0,1,2,\cdots [\ft{D-1}2]\,.
\end{eqnarray}
The procedure to obtain the explicit form of these generators in the
above coordinates were given in \cite{cls}.

We are looking for the spin-$\ell$ solution
$\psi^{(\ell)}_{\mu_1\cdots\mu_\ell}$ that is the highest-weight
state under the $SO(2,D-1)$ isometry group. It is defined by
\begin{eqnarray}
&&(H_0 - E^{(\ell)}_0)\psi^{(\ell)}_{\mu_1\cdots\mu_\ell}=0\,,\qquad
(H_i-h_i)\psi^{(\ell)}_{\mu_1\cdots\mu_\ell}=0 \qquad {\rm for}
\qquad i=1,2,\cdots\,,\cr 
&&E_{\vec \alpha_x} \psi^{(\ell)}_{\mu_1\cdots\mu_\ell}=0\,,\qquad
\hbox{for all simple roots $\vec \alpha_x$ and hence all positive
roots.}\label{geneom}
\end{eqnarray}
We use $\psi$ to denote the naturally-complex solutions, whilst
$\phi$ is the real or imaginary part of $\psi$.  The action of the
generators on the solution $\psi^{(\ell)}_{\mu_1\cdots\mu_\ell}$ is
as Lie derivative. The field $\psi^{(\ell)}_{\mu_1\cdots\mu_\ell}$
is totally symmetric, satisfying the transverse and traceless
condition, namely
\begin{equation}
g^{\mu_1\mu_2}\psi^{(\ell)}_{\mu_1\mu_2\cdots\mu_\ell}=0\,,\qquad
\nabla^{\mu_1} \psi^{(\ell)}_{\mu_1\mu_2\cdots\mu_\ell}=0\,.\label{ttcond}
\end{equation}
Note that the Casimir operator ${\cal E}$ is related to the
covariant Laplace operator $\Delta$ as follows
\begin{equation}
{\cal E} = \sum_i H_i H_i + \sum_{x} \ft12 |\alpha_x|^2
(E_{\alpha_x} E_{-\alpha_x} +E_{-\alpha_x} E_{\alpha_x}) = -
L^2\Delta\,,
\end{equation}
where the two sums are over all the Cartan and root generators
respectively.  The Laplace acting on the spin-$\ell$ field is
given by
\begin{eqnarray}
\Delta \psi_{\mu_1\cdots\mu_\ell} &=& - \Box \psi_{\mu_1\cdots
\mu_\ell} + \ell\,\nabla_\lambda \nabla_{(\mu_1}
\psi^\lambda{}_{\mu_2\cdots\mu_\ell)}\cr 
&=&-\Big(\Box + \fft{\ell(\ell+D-2)}{L^2}\Big)
\psi_{\mu_1\cdots\mu_\ell}\,.
\end{eqnarray}
Thus the solutions of (\ref{geneom}) must also satisfy the equation
(\ref{adseom}) for appropriate $M^{(\ell)}$.

We are now in the position to present the solutions of
(\ref{geneom}). {\it A priori},
$\psi^{(\ell)}_{\mu_1\cdots\mu_\ell}$ are in general functions of
all coordinates, with parameters $E^{(\ell)}_0$ and $h_i$. However,
for non-trivial solutions in $D\ge 4$, we find that
\begin{equation}
h_1=\ell\,,\qquad h_i=0\,,\qquad {\rm for}\qquad i\ge 2\,,
\end{equation}
(In $D=3$, $h_1=\pm \ell$, owing to the fact that $SO(2,2)\sim
SL(2,R)\times SL(2,R)$.) The reason for $h_1$ being singled out here
is due to the specific choice of the definition of positive and
simple roots that we adopt, as in \cite{cls}.  The scalar $\ell=0$
solution is given by
\begin{equation}
\psi^{(0)}=\Phi \equiv e^{-{\rm i} E^{(0)}_0 \tau}
(\cosh\rho)^{-E^{(0)}_0}\,, \label{s=0sol}
\end{equation}
which is independent of any spherical coordinate. Thus this $\ell=0$
solution is a singlet under the $SO(D-1)$ subgroup of $SO(2,D-1)$.
It should be pointed that there exist scalar solutions in AdS that
form infinite-dimensional representations \cite{bala}, which we
shall not consider in this paper.  For general higher spins in $D\ge
4$, we find a recursive relation between the spin-$(\ell+1)$ and the
spin-$\ell$ solutions
\begin{eqnarray}
\psi^{(\ell+1)}_{\mu_1\cdots\mu_{\ell}\tau} &=& s_\rho s_1s_2\cdots s_{n-1}\,
e^{-{\rm i} \phi_1}\,\Big(\psi^{(\ell)}_{\mu_1\cdots\mu_{\ell}}\Big)
\Big|_{E_0^{(\ell)}\rightarrow E_0^{(\ell+1)}}\,,\cr 
\psi^{(\ell+1)}_{\mu_1\cdots\mu_{\ell}\rho} &=& {\rm i}\,c^{-1}_\rho
s_1s_2\cdots s_{n-1}\, e^{-{\rm i}
\phi_1}\,\Big(\psi^{(\ell)}_{\mu_1\cdots\mu_{\ell}}\Big)
\Big|_{E_0^{(\ell)}\rightarrow E_0^{(\ell+1)}}\,,\cr 
\psi^{(\ell+1)}_{\mu_1\cdots\mu_{\ell}\theta_i} &=& {\rm i}\, s_\rho
s_1s_2\cdots s_{n-1}\,\fft{c_i}{s_i}\, e^{-{\rm i}
\phi_1}\,\Big(\psi^{(\ell)}_{\mu_1\cdots\mu_{\ell}}\Big)
\Big|_{E_0^{(\ell)}\rightarrow E_0^{(\ell+1)}}\,,\cr 
\psi^{(\ell+1)}_{\mu_1\cdots\mu_{\ell}\phi_1} &=&
s_\rho s_1s_2\cdots s_{n-1}\,
e^{-{\rm i} \phi_1}\,\Big(\psi^{(\ell)}_{\mu_1\cdots\mu_{\ell}}\Big)
\Big|_{E_0^{(\ell)}\rightarrow E_0^{(\ell+1)}}\,,\cr 
\psi^{(\ell+1)}_{\mu_1\cdots\mu_{\ell}\phi_i} &=& 0\,,\qquad {\rm for}\qquad
i=2,3,\cdots\,,\label{recursion}
\end{eqnarray}
where
\begin{equation}
s_\rho = \sinh\rho\,,\qquad c_\rho = \cosh\rho\,,\qquad
s_i=\sin\theta_i\,,\qquad c_i=\cos\theta_i\,.
\end{equation}
In $D=3$, there is an additional class of solutions with
$h_1=-\ell$. They are given by the above with $\phi_1\rightarrow
-\phi_1$, and the components with odd number of $\phi_1$ index
change the sign as well. Starting from the $\ell=0$ solution
(\ref{s=0sol}), it is straightforward to generate the solutions for
all $\ell$ using the recursion relation (\ref{recursion}).  The
solutions thus generated are necessarily totally symmetric. The
transverse and traceless conditions (\ref{ttcond}) are all
satisfied. It is easy to verify that the $\ell=1$ and $\ell=2$
solutions are indeed precisely those obtained in \cite{cls} by a
brutal-force calculation.  The spin-2 modes in $D=3$ and 4 were
obtained in \cite{Li:2008dq} and \cite{Bergshoeff:2011ri}
respectively.  The spin-3 solution in $D=3$ was also constructed in
\cite{Chen:2011vp}.  (Compare our solutions with the previous known
examples in $D=3$, we have $E_0=h+\bar h$ and $\ell = h-\bar h$.)
This recursive relation makes it straightforward to write up the
explicit solution for any spin-$\ell$ in any dimensions.  It is thus
unnecessary and also too complex to present the full results for
higher $\ell$ and $D$, owing to the proliferation of the components.
It is perhaps instructive just to present a particular component,
namely
\begin{equation}
\psi^{(\ell)}_{\tau\cdots\tau} = (s_\rho s_1\cdots s_{n-1})^\ell\,
e^{-{\rm i}\, (E_0^{(\ell)}\, \tau +\ell\, \phi_1)}
(\cosh\rho)^{-E_0^{(\ell)}}\,.\label{taucomp}
\end{equation}

The box and the Laplace actions on the solutions are given by
\begin{eqnarray}
\Box\psi^{(\ell)}_{\mu_1\cdots\mu_\ell} &=& \fft{(E^{(\ell)}_0)^2
- (D-1) E^{(\ell)}_0 - \ell}{L^2} \,\psi^{(\ell)}_{\mu_1\cdots\mu_\ell}
\,,\cr 
\Delta\psi^{(\ell)}_{\mu_1\cdots\mu_\ell} &=& -
\fft{(E^{(\ell)}_0)^2 - (D-1) E^{(\ell)}_0 + \ell(\ell+D-3)}{L^2}
\,\psi^{(\ell)}_{\mu_1\cdots\mu_\ell}\,.
\end{eqnarray}
Compare the first equation above to (\ref{adseom}), we have
\begin{equation}
(M^{(\ell)})^2 = \Big(E^{(\ell)}_0 + \ell -2\Big) \Big(E^{(\ell)}_0
- \ell-D+3\Big)\,.\label{mass1}
\end{equation}
It is perhaps more natural to write (\ref{adseom}) as
\begin{equation}
(\Box + \fft{\ell}{L^2} - \hat M^2)\psi^{(\ell)}=0\,,\qquad
\hat M^2 = \fft{(E_0^{(\ell)})^2 - (D-1)
E_0^{(\ell)}}{L^2}\,,\label{boxeom}
\end{equation}
where $\hat M$ does not depend on $\ell$ manifestly.

    Having obtained the solutions corresponding to the
highest-weight state of the spin-$\ell$ representation, we can use
the negative-root generators to act on the solution to obtain the
full multiplet.  In $D=4$, we generate $2\ell +1$ solutions with
$h=\ell, \ell-1,\cdots, -\ell+1, -\ell$.  The procedure becomes more
complicated in higher dimensions.  In figure 1, we give an example
of how the 16-dimensional spin-3 representation in $D=5$ is
generated from the highest-weight $(3,0)$ state.  Note that
$E_{-\vec\alpha_2}$ and $E_{-\vec\alpha_3}$ are the two negative-root conjugates of the simple roots associated with the $SO(4)$ subgroup of $SO(2,4)$. The explicit expressions for these generators
in our AdS coordinates can be found in \cite{cls}.  In general, the dimension for the
massive spin-$\ell$ representation in $D$ dimensions is given by
\begin{equation}
n=\fft{(2\ell +D-3)\, (\ell+D-4)!}{(D-3)!\, \ell!}\,.
\end{equation}
\medskip

\begin{center}
\includegraphics[scale=0.45]{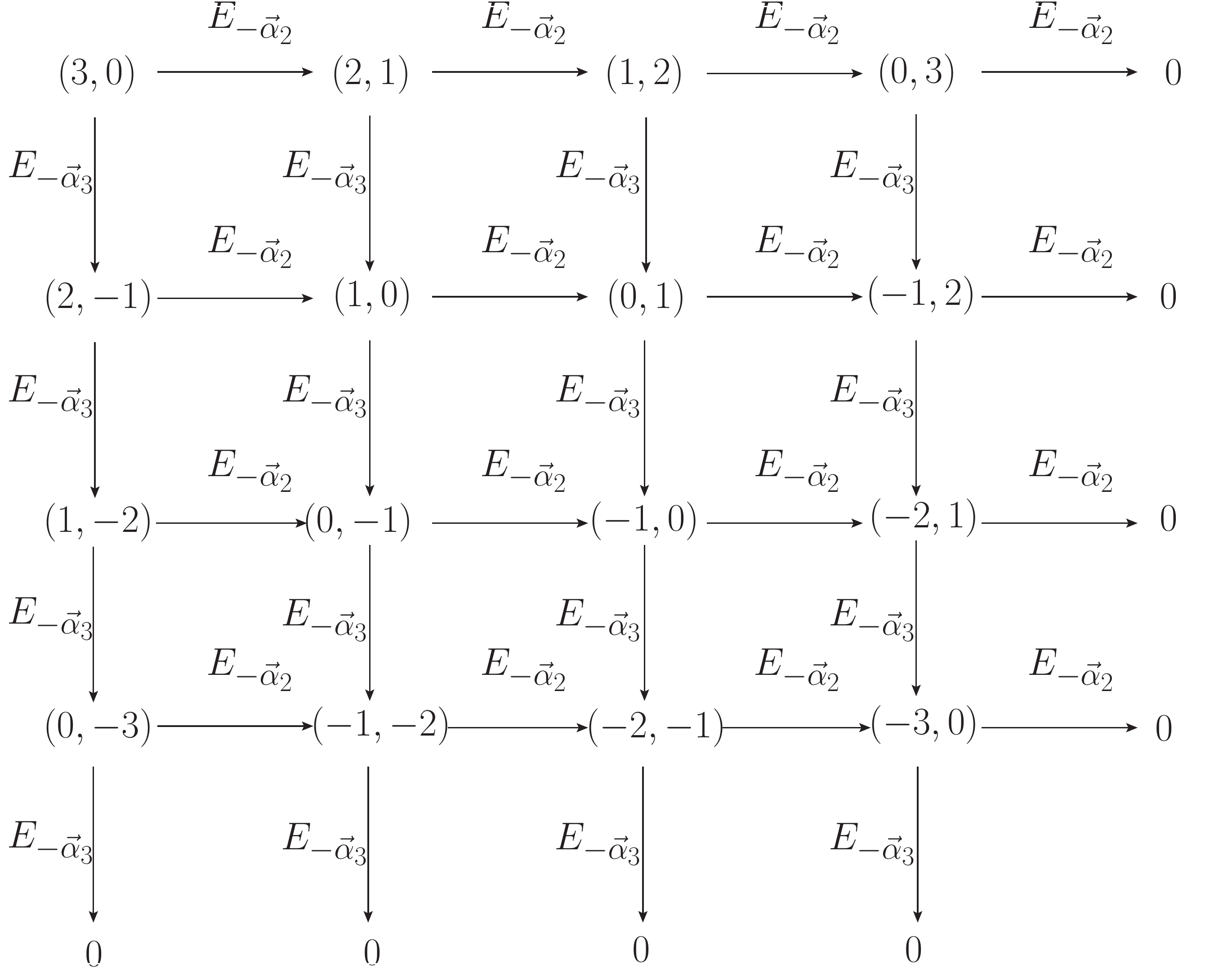}
\end{center}
Fig.~1: The generation of the 16-dimensional spin-$3$
representation of $SO(2,4)$ from the highest (3,0)-state.
\bigskip

    We now study the properties of our solutions.  For them
to be absent from exponential growth in time, it is necessary that
$E^{(\ell)}_0$ must be real.  It follows from (\ref{mass1}) that
\begin{equation}
(M^{(\ell)})^2 \ge - \fft{(2\ell + D-5)^2}{4 L^2}\,.\label{bf1}
\end{equation}
This is the generalized Breitenlohner-Freedman (BF) bound for
higher-spin fields.  The apparent difference of this bound for
$\ell=0$ with the usual BF bound is due to the different definitions
of the mass. Expressed in terms of $\hat M$, defined in
(\ref{boxeom}), we have
\begin{equation}
\hat M^2 \ge - \fft{(D-1)^2}{4 L^2}\,.\label{bf2}
\end{equation}

    Although higher spins in AdS backgrounds have the
analogous BF bounds, their falloff behavior with
negative mass square may differ from that of the scalars.
The leading term of the $\psi_{\tau\cdots\tau}$ (\ref{taucomp}) in
the AdS boundary is given by
\begin{equation}
\psi^{(\ell)}_{\tau\cdots\tau} \sim
\fft{1}{r^{E^{(\ell)}_0-\ell}}\,,\label{genfalloff}
\end{equation}
where $r=\sinh\rho\rightarrow \infty$.  The falloff of the AdS black
hole is given by $\delta g_{\tau\tau}\sim 1/r^{D-3}$. In \cite{cls}, the black hole falloff
is used to define the massless higher-spin modes.  It is easy to see
that indeed the condition
\begin{equation}
E^{(\ell)}_0-\ell = D-3
\end{equation}
corresponds precisely $M^{(\ell)}=0$ in (\ref{mass1}).  Thus for
solutions with $(M^{(\ell)})^2\ge 0$, it follows from (\ref{mass1})
that there must exist a branch that has the falloff faster than that of the black hole.

     In the asymptotically-flat space-time, the consistent boundary
condition is defined such that no mode should have a falloff slower than the $1/r^{D-3}$.  In solutions that is asymptotic to AdS, more general boundary conditions are allowed, which contain slower falloffs.  There is no unique choice for consistent boundary conditions.
The strongest and non-trivial boundary condition is that all modes should have the same or faster falloffs than $1/r^{D-3}$.  We call this strong boundary condition.

      For solutions that satisfy the bound (\ref{bf1}), but with
$(M^{(\ell)})^2<0$, two situations can arise.  The first is that
$D-5+2\ell\le 0$, the solutions satisfy the strong boundary condition. This include the scalar modes in $D=4$ and 5.  For $D-5+2\ell > 0$, the
solution has a slower falloff than what is required by the strong
boundary condition.  It is argued in \cite{Lu:2011ks} that the ghost massive graviton in extended gravities with such falloffs could be
truncated out by the strong AdS boundary condition.

      Finally, we would like to point out that in
higher-derivative theories, equations like $(\Box +\ell/L^2-\hat M^2)^2 \psi^{(\ell)}_{\mu_1\cdots\mu_\ell} =0$ may arise, in which case there exist log modes which are given by our solutions multiplied by a universal overall factor ${\it i}\,(\tau + \log(\cosh\rho))$.

To summarize, we consider free higher integer spins in AdS backgrounds in diverse dimensions.  We construct explicit
solutions corresponding to the highest-weight states of the
spin-$\ell$ representations of the $SO(2,D-1)$ isometry groups of
the AdS space-times.  Such a solution is the eigenstate of the
Cartan generators, annihilated by all positive-root generators. The
full multiplet can then be obtained by acting on
the solution repetitively with negative-root generators. These
explicit solutions enable us to obtain the generalized
Breitenlohner-Freedman bound for the higher spins, which allow
negative mass square. However, we find that the solutions with
negative mass square do not in general satisfy the strong AdS
boundary condition.  Our results should be useful in studying the
spectrum and the properties of the linearized modes in AdS gravities coupled to higher spins.

\appendix
\section*{Appendix: Gauge-fixing for massless higher-spin fields}

In this appendix, we examine the gauge fixing of the
gauge-invariant equation (\ref{adsmassless}).  The procedure does
not apply for $\ell=0$.  For $\ell=1$, it is given by the standard transverse condition.  For $\ell=2$, we adopt the following
gauge condition
\begin{equation}
\nabla^\mu \phi^{(2)}_{\mu\nu} = \nabla_\nu \phi^{(2)}\,,
\end{equation}
where $\phi^{(2)}\equiv g^{\mu\nu} \phi^{(2)}_{\mu\nu}$.  Note that
in this appendix, we use the notation
$\phi^{(\ell)}_{\mu_1\cdots\mu_k}$ with $k < \ell$ to denote the
field for which a trace or multiple traces are taken.  It is straightforward
to see that applying this gauge and taking the trace of
(\ref{adsmassless}), we have $\phi^{(2)}/L^2=0$, and we arrive at the
transverse and traceless conditions (\ref{tt}).  If the theory has
no cosmological constant, then this gauge is somewhat singular in
that there is no equation for the trace mode $\phi^{(2)}$ at all.
This gauge choice was first used in \cite{Li:2008dq} for studying chiral
gravity in three dimensions.  It was later adopted to study critical
phenomena in extended gravities in four \cite{lpcritical} and
higher dimensions \cite{dllpst}.

For $\ell=3$, we consider the following gauge condition
\begin{equation}
\nabla^\mu \phi^{(3)}_{\mu\nu\rho} = 2\nabla_{(\nu}
\phi^{(3)}_{\rho)}\,,
\end{equation}
which implies that $\nabla^\mu \phi^{(3)}_\mu =0$. Substituting
these into (\ref{adsmassless}) and taking a trace, we find that the
transverse and traceless conditions (\ref{tt}) come out naturally.

    The situation for $\ell\ge 4$ is different. Let us focus on
$\ell=4$.  The most general gauge condition is given by
\begin{equation}
\nabla^{\mu_1} \phi^{(4)}_{\mu_1\mu_2\mu_3\mu_4} = 3\alpha
\nabla_{(\mu_2} \phi^{(4)}_{\mu_3\mu_4)} + 3\beta\, g_{(\mu_2\mu_3}
\nabla_{\mu_4)} \phi^{(4)}\,,
\end{equation}
where $\alpha$ and $\beta$ are constants to be determined.
Substituting this into (\ref{adsmassless}), we find that the
derivative terms on $\phi^{(4)}$ cannot be canceled regardless the
choices of $\alpha$ and $\beta$, and that the equation for
$\phi^{(4)}$, which turns out to be independent of $(\alpha,\beta)$, is given by
\begin{equation}
\Big(\Box - \fft{2(D+1)}{L^2}\Big) \phi^{(4)}=0\,.
\end{equation}
It is of interest to note that the scalar trace mode $\phi^{(4)}$ appears
massive, compared to (\ref{adsmassless}).  It is thus more subtle to eliminate this mode, analogous to the subtlety in eliminating the trace mode in Einstein gravity in the De Donder gauge.

     We see that for $\ell=2,3$, the transverse and traceless
conditions can be derived from some proper choice of gauging. For
$\ell\ge 4$, the situation is more subtle, and we simply impose the conditions (\ref{tt}) by hand.

\section*{Acknowledgement}

We are grateful to Haishan Liu, Ergin Sezgin, Zhao-Long Wang for
useful discussions.  H.L.~is supported in part by the NSFC grant
11175269.  H.L.~is grateful to the Korea Institute for Advanced Study for hospitality during part of the work.


\begin{thebibliography}{99}

\bibitem{fp}
  M.~Fierz, W.~Pauli,
{\it On relativistic wave equations for particles of arbitrary spin
in an electromagnetic field,}
  Proc.\ Roy.\ Soc.\ Lond.\  {\bf A173}, 211-232 (1939).

\bibitem{sh}
  L.P.S.~Singh, C.R.~Hagen,
{\it Lagrangian formulation for arbitrary spin. 1. The boson case,}
  Phys.\ Rev.\  {\bf D9}, 898-909 (1974).

\bibitem{fron}
  C.~Fronsdal,
{\it Massless Fields with Integer Spin,}
  Phys.\ Rev.\  {\bf D18}, 3624 (1978).

\bibitem{ad}
  C.~Aragone and S.~Deser,
{\it Consistency problems of hypergravity,}
  Phys.\ Lett.\  B {\bf 86}, 161 (1979).

\bibitem{vasi1} M.A.~Vasiliev,
{\it Consistent equation for interacting gauge fields of all spins
in (3+1)-dimensions,}
  Phys.\ Lett.\  {\bf B243}, 378-382 (1990).

\bibitem{vasi2}
  M.A.~Vasiliev,
{\it Progress in higher spin gauge theories,} hep-th/0104246.

\bibitem{sun}
  B.~Sundborg,
{\it Stringy gravity, interacting tensionless strings and massless
higher spins,}
  Nucl.\ Phys.\ Proc.\ Suppl.\  {\bf 102}, 113-119 (2001).
  [hep-th/0103247].

\bibitem{dnw}
  L.~Dolan, C.~R.~Nappi, E.~Witten,
{\it Conformal operators for partially massless states,}
  JHEP {\bf 0110}, 016 (2001).
  [hep-th/0109096].

\bibitem{Polyakov:2001af}
  A.~M.~Polyakov,
{\it Gauge fields and space-time,}
  Int.\ J.\ Mod.\ Phys.\  {\bf A17S1}, 119-136 (2002).
  [hep-th/0110196].

\bibitem{ss1}
  E.~Sezgin, P.~Sundell,
{\it Massless higher spins and holography,}
  Nucl.\ Phys.\  {\bf B644}, 303-370 (2002).
  [hep-th/0205131].

\bibitem{ss2}
  E.~Sezgin, P.~Sundell,
{\it Analysis of higher spin field equations in four-dimensions,}
  JHEP {\bf 0207}, 055 (2002).
  [hep-th/0205132].

\bibitem{Girardello:2002pp}
  L.~Girardello, M.~Porrati, A.~Zaffaroni,
{\it 3-D interacting CFTs and generalized Higgs phenomenon in higher
spin theories on AdS,}
  Phys.\ Lett.\  {\bf B561}, 289-293 (2003).
  [hep-th/0212181].

\bibitem{dw}
  S.~Deser, A.~Waldron,
{\it Arbitrary spin representations in de Sitter from dS/CFT with
applications to dS supergravity,}
  Nucl.\ Phys.\  {\bf B662}, 379-392 (2003).
  [hep-th/0301068].

\bibitem{Bonelli:2003zu}
  G.~Bonelli,
{\it On the covariant quantization of tensionless bosonic strings in
AdS space-time,}
  JHEP {\bf 0311}, 028 (2003)
  [arXiv:hep-th/0309222].

\bibitem{Bianchi:2004xi}
  M.~Bianchi,
{\it Higher spins and stringy AdS$_5 \times S^5$,}
  Fortsch.\ Phys.\  {\bf 53}, 665-691 (2005).
  [hep-th/0409304].

\bibitem{Sorokin:2004ie}
  D.~Sorokin,
{\it Introduction to the classical theory of higher spins,}
  AIP Conf.\ Proc.\  {\bf 767}, 172-202 (2005).
  [hep-th/0405069].

\bibitem{Bouatta:2004kk}
  N.~Bouatta, G.~Compere, A.~Sagnotti,
{\it An introduction to free higher-spin fields,} [hep-th/0409068].

\bibitem{Vasiliev:2004cp}
  M.A.~Vasiliev,
  {\it Higher spin gauge theories in any dimension,}
  Comptes Rendus Physique {\bf 5}, 1101-1109 (2004).
  [hep-th/0409260].

\bibitem{Bekaert:2005vh}
  X.~Bekaert, S.~Cnockaert, C.~Iazeolla, M.~A.~Vasiliev,
{\it Nonlinear higher spin theories in various dimensions,}
 [hep-th/0503128].

\bibitem{Bianchi:2005yh}
  M.~Bianchi, V.~Didenko,
{\it Massive higher spin multiplets and holography,}
  [hep-th/0502220].

\bibitem{Bianchi:2005ze}
  M.~Bianchi, P.~J.~Heslop, F.~Riccioni,
{\it More on La Grande Bouffe,}
  JHEP {\bf 0508}, 088 (2005).
  [hep-th/0504156].

\bibitem{Li:2008dq}
  W.~Li, W.~Song, A.~Strominger,
{\it Chiral gravity in three dimensions,}
  JHEP {\bf 0804}, 082 (2008).
  [arXiv:0801.4566 [hep-th]].

\bibitem{Bergshoeff:2011ri}
  E.A.~Bergshoeff, O.~Hohm, J.~Rosseel, P.K.~Townsend,
{\it Modes of log gravity,}
  Phys.\ Rev.\  {\bf D83}, 104038 (2011).
  [arXiv:1102.4091 [hep-th]].

\bibitem{cls}
  Y.X.~Chen, H.~L\"u and K.N.~Shao,
{\it Linearized modes in extended and critical gravities,}
  arXiv:1108.5184 [hep-th].

\bibitem{bala}
  V.~Balasubramanian, P.~Kraus and A.E.~Lawrence,
{\it Bulk versus boundary dynamics in anti-de Sitter space-time,}
  Phys.\ Rev.\  D {\bf 59}, 046003 (1999)
  [arXiv:hep-th/9805171].

\bibitem{Chen:2011vp}
  B.~Chen, J.~Long and J.b.~Wu,
{\it Spin-3 topological massive gravity,}
  arXiv:1106.5141 [hep-th].

\bibitem{Lu:2011ks}
  H.~L\"u, Y.~Pang, C.N.~Pope,
{\it Conformal gravity and extensions of critical gravity,}
  Phys.\ Rev.\  {\bf D84}, 064001 (2011).
  [arXiv:1106.4657 [hep-th]].

\bibitem{lpcritical}
  H.~L\"u, C.N.~Pope,
{\it Critical gravity in four Dimensions,}
  Phys.\ Rev.\ Lett.\  {\bf 106}, 181302 (2011).
  [arXiv:1101.1971 [hep-th]].

\bibitem{dllpst}
  S.~Deser, H.~Liu, H.~L\"u, C.N.~Pope, T.C.~Sisman, B.~Tekin,
{\it Critical points of $D$-dimensional extended gravities,}
  Phys.\ Rev.\  {\bf D83}, 061502 (2011).
  [arXiv:1101.4009 [hep-th]].

\end{thebibliography}
\end{document}